\documentclass[a4paper,fleqn,usenatbib]{mnras}

\usepackage{mathptmx}

\usepackage[T1]{fontenc}
\usepackage{ae,aecompl}

\usepackage{graphicx}	
\usepackage{amsmath}	
\usepackage{amssymb}	

\usepackage{amsfonts}
\usepackage{natbib}
\usepackage{nth}

\def\aj{AJ}                  
\def\apj{ApJ}                  
\def\apjl{ApJL}    
\def\mnras{MNRAS}                  
\def\araa{ARA\& A}
\def\pasp{PASP}
\def\apjs{ApJS}  
\def\aap{A\&A}  
  
\def\apss{Ap\&SS}
\def\nat{Nature}

\newcommand{\nii}{[N{\small II}]}
\newcommand{\sii}{[S{\small II}]}
\newcommand{\oiii}{[O{\small III}]}
\newcommand{\oi}{[O{\small I}]}

\title[SAMI: disk winds with radio emission]{The SAMI Galaxy Survey: Disk-halo interactions in radio-selected star-forming galaxies}

\author[S. K. Leslie et al.]{
S. K. Leslie,$^{1,2,3}$\thanks{E-mail: leslie@mpia.de}
J. J. Bryant,$^{2,4,5}$
I.-T. Ho,$^{1,3,6}$
E. M. Sadler,$^{2,4}$
A. M. Medling,$^{1,7}$\thanks{Hubble Fellow}
\newauthor{
B. Groves,$^{1}$
L. J. Kewley,$^{1}$
J. Bland-Hawthorn,$^{4}$
S. M. Croom,$^{2,4}$
O. I. Wong,$^{8}$
}
\newauthor{
S. Brough,$^{5}$
E. Tescari,$^{2,9}$
S. M. Sweet,$^{1}$
R. Sharp,$^{1}$
A. W. Green,$^{5}$ 
}
\newauthor{
\'{A}. R. L\'{o}pez-S\'{a}nchez,$^{5,10}$
J. T. Allen,$^{2,4}$
L. M. R. Fogarty,$^{4}$
M. Goodwin,$^{5}$
}
\newauthor{
J. S. Lawrence,$^{5}$
I.S. Konstantopoulos,$^{5,11}$
M. S. Owers,$^{5,10}$
S. N. Richards$^{2,4,5}$
}
\\
$^{1}$Research School of Astronomy and Astrophysics, Australian National University, Canberra, ACT 2611, Australia\\
$^{2}$ARC Centre of Excellence for All-sky Astrophysics (CAASTRO)\\
$^{3}$ Max-Planck-Institut f\"{u}r Astronomie, K\"{o}nigstuhl 17, 69117 Heidelberg, Germany\\
$^{4}$Sydney Institute for Astronomy (SIfA), School of Physics, The University of Sydney, NSW 2006, Australia\\
$^{5}$The Australian Astronomical Observatory, 105 Delhi Rd, North Ryde, NSW 2113, Australia\\
$^{6}$Institute for Astronomy, University of Hawaii, 2680 Woodlawn Drive, Honolulu, HI 96822, USA\\
$^{7}$Department of Astronomy, California Institute of Technology, 1200 E California Blvd, Pasadena, CA 91125, USA\\
$^{8}$ICRAR-M468, University of Western Australia 35 Stirling Highway, Crawley WA 6009, Australia\\
$^{9}$School of Physics, University of Melbourne Parkville, VIC 3010 Australia\\
$^{10}$Department of Physics and Astronomy Macquarie University NSW 2109 Australia\\
$^{11}$Envizi Suite 213, National Innovation Centre, Australian Technology Park, 4 Cornwallis Street, Eveleigh NSW 2015, Australia
}

\date{Accepted 2017 June 30. Received 2017 June 2; in original form 2016 December 13}
\pubyear{2017}

\begin{document}
\label{firstpage}
\pagerange{\pageref{firstpage}--\pageref{lastpage}}
\maketitle

\begin{abstract}

In this paper, we compare the radio emission at 1.4 GHz with optical outflow signatures of edge-on galaxies.
We report observations of six edge-on star-forming galaxies in the Sydney-AAO Multi-object Integral-field spectrograph (SAMI) Galaxy Survey with 1.4 GHz luminosities $> 1\times10^{21}$ W Hz$^{-1}$.
Extended minor axis optical emission is detected with enhanced \nii/H$\alpha$ line ratios and velocity dispersions consistent with galactic winds in three of six galaxies. These galaxies may host outflows driven by a combination of thermal and cosmic ray processes.
We find that galaxies with the strongest wind signatures have extended radio morphologies. Our results form a baseline for understanding the driving mechanisms of galactic winds. 

\end{abstract}

\begin{keywords}
ISM: jets and outflows, ISM: cosmic rays, galaxies: evolution, galaxies: kinematics and dynamics
\end{keywords}

\section{Introduction}
Galactic winds, driven by star formation or active galactic nuclei (AGN), are arguably the most important feedback processes in galaxy formation and evolution \citep{veilleux05}. Without incorporating outflows, galaxy formation models predict galaxy star formation rates (SFRs) much higher than those observed; large-scale outflows remove gas content and therefore suppress the SFRs of galaxies (e.g. \citealt{cole00,springel03,keres09} and references therein). Outflows from galaxies also drive metal-enriched gas out into the surrounding medium and are therefore thought to be responsible for the heavy elements present in the circumgalactic medium \citep{tumlinson11,werk13}. 
Galactic winds are ubiquitous in local galaxies with a high SFR surface density ($>0.1$M$_\odot$ yr$^{-1}$ kpc$^{-2}$), with outflow velocities seemingly correlated with specific SFR ($\dot{\text{M}}/\text{M}_*$; \citealt{heckman90,grimes09}) or circular velocity (e.g. \citealt{heckman15}). The discovery of powerful winds in high-redshift galaxies (e.g. \citealt{pettini00,steidel10,genzel11,newman12,	genzel14}) further implicates wind-related feedback processes as key to the chemical and thermal evolution of galaxies and the intergalactic medium (IGM). 

The galactic fountain model of \cite{shapiro76} describes how outflows with velocities less than the escape velocity will eventually fall back onto the galactic disk, resulting in the circulation of matter through disk-halo interactions. Disk-halo interactions include outflows from star forming regions \citep{normandeau96}, and cool clouds falling back into the disk from the halo \citep{pietz96}. The chimney model of \cite{norman89} suggests more localised outflows than the fountain model. 
Neutral hydrogen super-shells are driven by stellar winds and supernova activity and leave relative `holes' where the stars were. If the super-shells are sufficiently energetic, they can break through the gas layers of the disk, resulting in a chimney shaped structure. Such chimneys would provide an unobstructed route for photons to escape the disk and ionize gas at large scale heights. Magnetic fields would also follow these chimneys or filaments, allowing cosmic rays (CRs) to be more easily advected into the halo. 

CRs provide an important additional pressure on the gas, thereby helping to drive gas into the halo. Recent simulation papers such as \cite{uhlig12}, \cite{booth13}, \cite{salem14}, \cite{girichidis16}, \cite{simpson16}, \cite{ruszkowski16} and \cite{weiner16} explore the importance of CRs in driving winds in star-forming galaxies.
CR-driven winds are able to explain outflows with a low velocity ($\sim$20 km s$^{-1}$) and correspondingly low SFR surface densities (below 0.1 M$_\odot$yr$^{-1}$kpc$^{-2}$). CR energy is not dissipated as fast as thermal energy. CR pressure is able to act over larger distances than thermal pressure, thus helping to drive winds where a thermal wind alone would fail \citep{everett08}. On the other hand, CR pressure is relatively inefficient at driving winds from the galactic mid-plane compared to gas pressure when the wind is sub-Alfv\'{e}nic \citep{everett08}. Thus, a mixture of thermal pressure, to efficiently launch the wind and CR pressure, to efficiently drive the wind to large radii, would be desirable for expelling gas from a galaxy.
 
Edge-on galaxies provide the best view of extraplanar gas at the interface between a galaxy disk and halo. Observations of nearby edge-on galaxies show that gas, dust and CRs exist above and below the disk \citep{beuermann85, madsen05, irwin06, popescu04}. H$\alpha$ narrow band imaging has revealed strong starburst-driven winds in galaxies such as M82 \citep{axon78,bland88} and NGC 253 \citep{westmoquette11}, which have bipolar structures centred on a central starburst and extending several kiloparsecs in disk height. Studies have also revealed that extended diffuse emission and filamentary structures are common in nearby galaxies: More than $50$\% of galaxies have extraplanar diffuse emissions from 1-2 kpc to even 4 kpc above the disk \citep{rossa00,rossa03, miller03}.  However, halos are intrinsically faint (low surface brightness) emitters, rendering them hard to detect (e.g. \citealt{hummel91}). 

Radio continuum emission that extends along the minor axis of edge-on spiral galaxies to form a ``thick disk'' has been detected in nearby galaxies such as NGC 4565, NGC 5909 \citep{hummel83}, NGC  5775 \citep{duric98}, NGC 3044 \citep{sorathia94} and NGC 3556 \citep{bloemen93}. \cite{krause11} measured the vertical scale heights of the radio emission at 6 cm for five edge-on galaxies and found they all had similar vertical scale heights for the thin and thick disks, with a mean thick-disk scale height of 1.8$\pm0.2$ kpc. X-shaped magnetic fields in the thick disk/halo were also detected for 8 edge-on galaxies spanning a range of Hubble types and SFRs 0.6$\leq SFR \leq$7.3 M$_\odot$ yr$^{-1}$ \citep{krause11}. The strength of the X-shaped fields can be explained by models including a galactic wind \citep{krause11}. Biconical outflows were also predicted by the CR-driven models of \cite{salem14}. 

 At optical wavelengths, high resolution spectroscopy can be used to identify shocks from galactic winds in galaxies \citep{veilleux05}. Emission-line ratios typical of shock excitation (e.g. \nii /H$\alpha>1$) indicate winds in the absence of AGN activity. Studies of local galaxies show that outflows are usually associated with interstellar shocks that excite optical line emission, enhancing line ratios such as \nii /H$\alpha$, \sii /H$\alpha$  and \oi /H$\alpha$ \citep{rich10,rich11,soto12,sotomartin12}. For inclined galaxies, one can use the excitation contrast in the shock-excited wind material and the star-forming disk to search for and detect galactic winds (e.g. \citealt{veilleux02}). Samples of a few tens of galactic wind galaxies have been studied using slit spectroscopy and narrow-band imaging (e.g., \citealt{heckman90,veilleux03,veilleux05}).

Integral field spectroscopy (IFS) is an ideal tool for investigating galactic winds, allowing spatially resolved kinematic and excitation information to be obtained simultaneously. The use of optical diagnostic diagrams with spatially resolved data can allow the separation of winds driven by starbursts from those driven by AGN \citep{sharp10}. Therefore, IFS of highly-inclined galaxies is an excellent method for identifying wind signatures in star-forming galaxies and studying disk-halo interactions in more detail.

We draw optical IFS data from an on-going multiplexed integral field unit (IFU) survey: the Sydney-AAO Multi-object Integral-field spectrograph (SAMI) Galaxy Survey \citep{croom12,bryant15}. 
The ability of SAMI to detect galactic winds in star-forming galaxies has already been demonstrated by \cite{fogarty12} and \cite{ho14}. The high spectral resolution of SAMI (R$\approx$4500 in the red arm) allowed \cite{ho14} to decompose the spectral lines from a moderately inclined disk galaxy (i=43$^\circ$) into three separate kinematic components having different velocities, velocity dispersions and line ratios; one of the components traced a bipolar outflow. A sample of 40 star-forming edge-on galaxies in the SAMI survey was studied by \cite{ho16}. \cite{ho16} found large-scale winds preferentially in galaxies with high star formation rate (SFR) surface densities and recent bursts of star formation.

Low frequency (1.4 GHz) radio observations are sensitive to the synchrotron emission resulting from CR electrons in the presence of a magnetic field.
Using radio images from the Faint Images of the Radio Sky at Twenty-centimetres (FIRST) survey, we can test which conditions enable CRs to escape to the largest disk heights. Following the work of \cite{ho16}, we discuss six edge-on star-forming galaxies in the SAMI Galaxy Survey that show evidence for low velocity outflows and have strong radio continuum emission. 

The paper is structured as follows. In Section \ref{sample} we discuss our sample selection, detail our observations, data reduction, and spectral analysis. We analyse emission line ratio maps, velocity and velocity dispersion maps in Section \ref{analysis}. We also employ the gas kinematic classification developed in \cite{ho16} to identify wind-dominated galaxies. In Section 4 we present the FIRST maps of the galaxies and discuss the extent of the radio emission and the location of our sample with respect to the radio-far-infrared relation. 
We discuss our results in the context of disk-halo interactions in Section \ref{discussion}. Finally, we give our conclusions in Section 6.

Throughout this paper we assume a (h,$\Omega_m, \Omega_\Lambda$) = (0.7,0.3,0.7) Cosmology and a Salpeter initial mass function.

\section{Sample}\label{sample}

\subsection{SAMI Galaxy Survey data}
To investigate disk-halo interactions in star-forming galaxies, we identify edge-on galaxies observed by the SAMI Galaxy Survey that have measured radio continuum emission.
The SAMI Galaxy Survey covers a broad range in stellar mass and environment. Full details of the target selection can be found in \cite{bryant15}. 
Briefly, the SAMI Galaxy Survey includes four volume-limited samples based on pseudo-stellar mass
cuts all selected from the Galaxy
and Mass Assembly (GAMA) project \citep{driver11}. To extend the survey sample to high density environments, 8 clusters with virial masses $>10^{14}$ M$_\odot$ are also targeted by the survey \citep{owers17}.
SAMI-GAMA targets cover 0.004 $<$ z $<$ 0.095, Petrosian r-band magnitudes $<$ 19.8, and a stellar mass range of
$10^7-10^{12}$ M$_\odot$. The GAMA regions were selected because of the deep spectroscopy down to r$<$ 19.8 mag of $\sim$300 000 galaxies, and the
vast array of ancillary data available, including $ugrizYJHK$ images through to radio bands.

IFS observations were performed using SAMI \citep{croom12} and are described in Table \ref{observations}. SAMI is mounted at the prime focus on the Anglo-Australian Telescope that provides a 1 degree diameter field of view. SAMI uses 13 fused fibre bundles (hexabundles, \citealt{bryant14,blandhawthorn11,bryant11}) with a high (75\%) filling factor. Each bundle contains 61 fibres of 1.6 arcsec diameter resulting in each IFU having a diameter of 15 arcsec. The IFUs, as well as 26 sky fibres, are plugged into pre-drilled plates using magnetic connectors. SAMI fibres are fed to the double-beam AAOmega spectrograph \citep{sharp06}. The SAMI Galaxy Survey uses the 570V grating at 3700-5700\AA~ giving a resolution of R=1730 ($\sigma$=74 km s$^{-1}$), and the 1000R grating from 6250-7350\AA~ giving a resolution of R=4500 ($\sigma$=29 km s$^{-1}$).

The data were reduced using the SAMI data reduction pipeline described in \cite{sharp15}, \cite{allen15}, and Green et al. (in prep).
Final SAMI data products are sampled to a grid with square pixels 0.5 arcsec wide.

The seeing conditions of our observations are measured from reduced data cubes of a calibration star observed simultaneously with the science targets. As described in \cite{allen15}, the spatial point spread function (PSF) of the SAMI data cubes is primarily determined by the atmospheric conditions at the AAT. A 2D Moffat distribution was fit to the red cube of the calibration stars to provide an estimate of the seeing disk. We note that the PSF profile is a function of wavelength, with larger seeing disks on average towards the blue. We report the FWHM of the resulting fits in Table \ref{observations}.

\subsection{FIRST radio data}
To investigate the relationship between CR electrons and their host galaxies, we select SAMI targets that have strong 1.4 GHz radio continuum emission. Synchrotron emission from the CR electrons in a galaxy will dominate the radio emission of star forming galaxies at frequencies lower than $\sim$30 GHz \citep{condon92}. The non-thermal synchrotron component comprises $\sim90\%$ of the radio emission at 1.4 GHz for star forming galaxies \citep{condon92}.
Cosmic rays can travel 1 to 3 kpc during their lifetime in a galaxy disk or filament respectively \citep{collins00}. Detecting CRs at heights $>3$ kpc above a galaxy disk could be an indication of a galactic wind or chimney.\footnote{The presence of a radio-jet can also result in radio emission $>3$ kpc away from a galaxy disk. Jets and galactic winds can be differentiated via their radio morphology.}
We draw radio continuum data from the 
Faint Images of the Radio Sky at Twenty-centimetres (FIRST; \citealt{becker95}) survey, a 1.4 GHz continuum survey carried out by the National Radio Astronomy Observatory (NRAO) with the Very Large Array (VLA).
The FIRST survey was conducted in the B-configuration of the VLA, and thus has a resolution of $\sim$5.4 arcsec. At the 1 mJy source detection threshold, the survey is 80\% complete and has $\sim$90 sources per square degree, $\sim$35\% of which have resolved structure on scales from 2-30 arcsec. A publicly available source catalogue\footnote{ http://sundog.stsci.edu/first/catalogs.html} includes peak and integrated flux densities and sizes derived from fitting a two-dimensional Gaussian profile to each source. The astrometric uncertainty of the sources is 1 arcsec. 

\subsection{Cross-matched sample}
We positionally cross-matched potential SAMI targets with radio surveys FIRST and NVSS (NRAO VLA Sky Survey; \citealt{condon98}). Radio contours were overlaid on optical images to visually confirm matches. 
We find that $\sim$4\% of the 5546 potential SAMI targets in the GAMA regions have associated radio sources in either survey, with approximately two thirds of these being optically classified as star forming using the \cite{kewley01} classification and GAMA single fibre spectroscopy \citep{hopkins13}.

Due to sensitivity limits, FIRST primarily detects luminous galaxies with radio-SFRs $\gtrsim$ 1 M$_\odot$ yr$^{-1}$, that corresponds to galaxies more massive than 10$^{9.5}$M$_\odot$. The FIRST catalogue only contains galaxies with a high surface brightness, likely biasing our radio-selected sample towards a larger fraction of wind-dominated galaxies, because wind activity is thought to correlate with SFR surface density \citep{ho16,heckman90}.


\subsection{Selection criteria} 
Galaxies dominated by AGN were removed from the sample using the GAMA spectra \citep{liske15} or central SAMI spectra which probes the central 2 arcsec of the galaxies and the \cite{kewley06} classification scheme. However, some radio-AGN do not show strong optical AGN signatures \citep{sadler02,ivezic02}, so the removal of optically-luminous AGN does not necessarily mean that our sample is free of radio AGN. 
Thirty star forming galaxies, observed with SAMI as of June 2016, remain after removing AGN-host galaxies. 

For our analysis of disk-halo interactions, we make cuts on galaxy size and inclination.
Effective radii (r$_e$) and ellipticity were measured from S\'{e}rsic profile fits to reprocessed SDSS (Sloan Digital Sky Survey; \citealt{york00,stoughton02}) $r$ band images by the GAMA team \citep{kelvin12}.
To observe gas excitation above and below the disk, we require the galaxy effective radius to be less than the size of the SAMI IFUs, 15 arcsec (this excludes one galaxy, GAMA 204799). 
Galaxy inclination is determined using the galaxy minor-to-major axis ratio, $b/a$. We apply a cut to only include edge-on galaxies using the criteria $\frac{b}{a} \leq 0.35$. This corresponds to an inclination angle $i$ = 73$^\circ$ calculated using the \cite{hubble26} formula
$$\cos^2(i)=\frac{(\frac{b}{a})^2-q_0^2}{1-q_0^2},$$ where $q_0=0.2$ is related to the intrinsic flattening of the distribution of light of galacic spheroids. Additionally, GAMA 504713 was excluded from the analysis due to the poor fit of a single-S\'{e}rsic profile to the $r$-band photometry ($\chi^2\approx 2$).

\subsection{The Sample}

Our selection resulted in a sample of six edge-on, star-forming, radio continuum detected galaxies for which CR pressure could play an important role in driving disk-halo interactions. 
The six nearby edge-on galaxies in our sample, with GAMA catalogue identifications 551202, 417678, 593680, 600030, 376293, and 227607 (which we will refer to hereafter as Galaxies 1 to 6 respectively) are shown in Figure \ref{maps}. Row A contains SDSS \textit{gri} composite
images with the SAMI hexabundle field of view (15 arcsec in diameter) overlaid. 
The dust lanes and stellar bulges of Galaxies 1, 3, 4, 5, and 6 can be seen in the optical images.

\begin{table*}

\caption{Information from the GAMA second public data release and on the SAMI observations of the six galaxies in our sample.}\label{observations}
\begin{tabular}{|cc|ccccc|cc|cc|}
\hline
ID & GAMA ID & RA & Dec & z & $r_e^{(1)}$ & $i^{(2)}$ &log($\frac{M_*}{M_\odot}$)$^{(3)}$&SFR$^{(4)}$& Date Obs $^{(5)}$ &FWHM$^{(6)}$\\
 & & & & &arcsec& $^o$ &  & M$_\odot$ yr$^{-1}$ & & arcsec\\
\hline
1& 551202 & 09:17:28.99 & -00:37:14.1 & 0.0165 & 12.69&78.5 &9.967 & 4.59 & 04 Apr 2014 &  1.85\\
2&417678 & 08:50:57.17 & +02:20:46.2 & 0.0405 & 4.33 &90.0 &10.11 & 10.13 & 06 Apr 2014 & 2.29 \\
3&593680 & 14:29:46.05 & -00:09:08.6 & 0.0300 & 11.53&90.0 &10.411& 1.90 & 14 \& 16 Apr 2013& 2.18\\
4& 600030 & 08:53:58.68 & +00:21:00.1 & 0.0292 & 8.27&73.0 &10.237 & 5.10 & 13 Mar 2013&1.60 \\
5& 376293 & 08:51:11.41 & +01:30:06.1 & 0.0605 & 6.96 &82.2 &10.65 & 2.25 & 25 Jan 2015 & 2.72\\
6&227607 & 14:17:17.86 & +01:09:22.4 & 0.0545 & 10.69&76.0 &10.75 & 2.30 & 23 May 2015 &1.61\\
\hline
\end{tabular}
$^{(1)}$Effective radii and $^{(2)}$inclination angle from the $r$ band photometry \citep{kelvin12}. $^{(3)}$ GAMA stellar masses (M$_*$/M$_\odot$) \citep{taylor11}. $^{(4)}$ GAMA H$\alpha$ Star-formation rates \citep{hopkins13}. $^{(5)}$ The date(s) the target was observed with SAMI. $^{(6)}$The effective seeing of the SAMI observation.
\end{table*}

\subsection{Spectral analysis}\label{specanal}
We use the spectral fitting pipeline \textsc{lzifu} \citep{ho16b} to extract the emission line fluxes and kinematic information from the SAMI data cubes for each galaxy.  On a spaxel-by-spaxel basis, the continuum is modelled and subtracted and the emission lines are fit to the resulting spectrum. \textsc{lzifu} uses the penalized-pixel fitting routine (\textsc{ppxf}; \citealt{cappellari04}) to model the continuum spectrum using spectral synthesis models. We use templates from \cite{gonzalezdelgado05} made with Padova stellar evolutionary tracks, three metallicities, Z=0.004, 0.008 and 0.019 (solar), and 24 ages ranging from 0.004 to 11.220 Gyr.
Once the continuum is subtracted, we fit 11 strong optical lines simultaneously: [O{\footnotesize II}] 3726,29, H$\beta$, \oiii~4959,5007, [OI]6300, \nii~6548,84, H$\alpha$ and \sii~6716,31. Emission lines are fit with a single Gaussian component at each spaxel using the \textsc{mpfit} package, which performs a least-squares analysis using the Levenberg-Marquardt algorithm \citep{markwardt09}. The ability of \textsc{lzifu} to fit multiple Gaussian components to the emission line spectra is most useful where any outflowing component would be superimposed along our line of sight with the disk, such as for a face-on galaxy. For our edge-on sample, a one-component fit was preferred to a two component Gaussian fit based on the F-test and confirmed through visual inspection in the extraplanar regions. We thus use \textsc{lzifu} to only fit a single Gaussian per emission line. Spaxels located outside of 1 effective radius have mean reduced $\chi^2$ $\leq 1$ for all six galaxies ( reduced $\chi^2 =$ 0.919, 1.11, 0.905, 0.935, 0.965, 0.963 for galaxies 1 to 6 respectively). 
By fitting all lines simultaneously, every line is constrained to share the same velocity and velocity dispersion.

The H$\alpha$ flux maps of all six galaxies are shown in row B of Figure \ref{maps}. The H$\alpha$ emission traces the recent ($<$10 Myr) star formation activity (but it also suffers from dust extinction).

\section{Optical signatures of disk-halo interactions}\label{analysis}
In this section, we present our analysis of the emission line and gas kinematic properties of our galaxies. 

\subsection{Emission line ratio maps}
\begin{figure*}
\centering

\includegraphics[width=\linewidth]{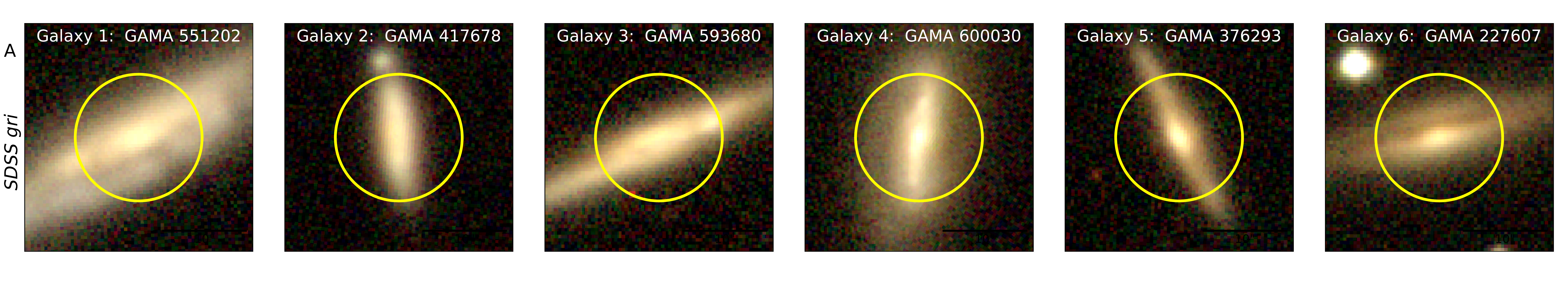}
\includegraphics[width=\linewidth]{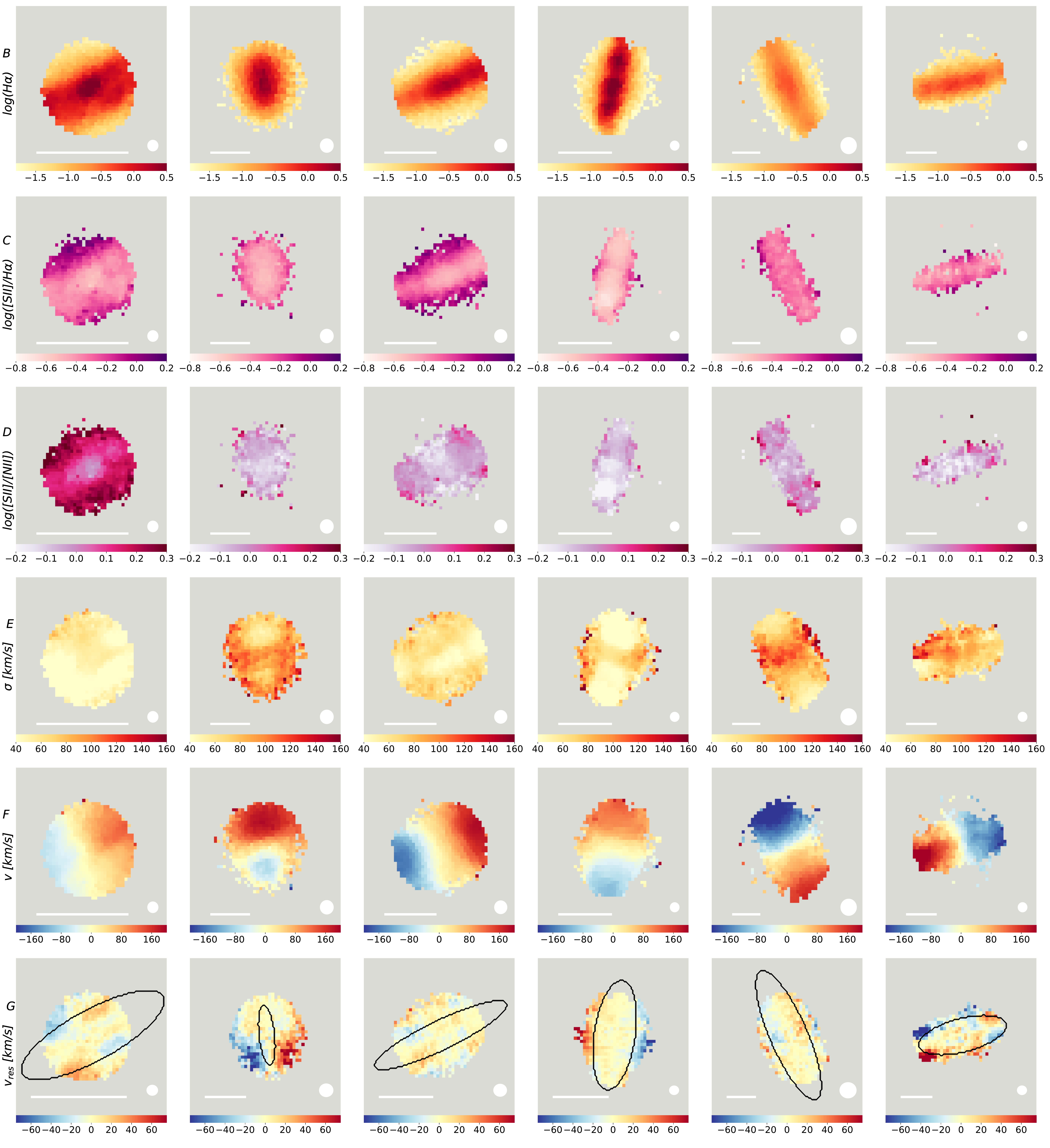}

\caption{Each column represents the properties of one galaxy. Row A shows 1$'\times$1$'$ Sloan Digital Sky Survey \textit{g r i} composite images with the SAMI hexabundle field of view (15 arcsec in diameter) overlaid.  Row B shows the log of the total H$\alpha$ flux (in units of 10$^{-16}$erg s$^{-1}$cm$^{-2}$). Row C shows maps of log(\sii /H$\alpha$). In all six galaxies, the line ratios increase with distance from the disk. 
Row D shows maps of log(\sii/\nii).
Row E shows maps of velocity dispersion, $\sigma$ (in km s$^{-1}$).
Row F shows velocity maps (in km s$^{-1}$). All galaxies show clear signs of rotation.
Row G shows velocity residual maps, obtained by reflecting the velocity field about the major axes and subtracting the reflected map from the original map shown in row F. Black ellipses represent 1 effective radius padded by 1 arcsecond, $\tilde{r_e}$).
In all panels, only spaxels with $S/N \geq 5$ in relevant properties are shown. White bars at the bottom of the maps of SAMI data indicate 5 kpc in projection at the redshift of each galaxy. The effective seeing disk of each observation is represented by a white circle at the bottom right of the maps. }\label{maps}
\end{figure*}

We examine maps of the \sii$\lambda$6716,31 doublet (hereafter \sii ) to H$\alpha$ flux ratio in row C of Figure
\ref{maps} as a first step towards understanding the processes at work in the galaxies.
The \sii /H$\alpha$ ratio is sensitive to metallicity and ionization
parameter \citep{kewley02,denicolo02,pettini04,kewley08} and has the advantage of not requiring reddening corrections. The \sii /H$\alpha$ ratio can be excited by interstellar shocks (e.g. \citealt{shull79,farage10,rich10}).
All six galaxies show the same trend: above and below the disk, the \sii /H$\alpha$ ratio increases.
Higher ratios in the outer regions usually correspond to shocked regions, as seen by \citealt{veilleux02}, \citealt{rich10}, \citealt{monrealibero06}, \citealt{sharp10}, 
\citealt{rich11}, and \citealt{fogarty12}. 

The \nii /H$\alpha$  line ratios, and \oi /H$\alpha$ line ratio (see Figures \ref{lvdisp} and \ref{BPTs}) also follow the same trend as the \sii /H$\alpha$ maps, with line ratios increasing with distance from the disk for all six galaxies.

Spiral galaxies with a layer of vertically extended diffuse ionised gas (eDIG) also show a rise in \nii /H$\alpha$ and \sii /H$\alpha$. Diffuse ionized gas is a warm (10$^4$K), low density (10$^{-1}$cm$^{-3}$), ionized medium, known in our own galaxy as the Reynolds layer \citep{reynolds93}. There is a lack of consensus in the literature on how to explain the emission properties of eDIG \citep{domgorgen94,veilleux95, blandhawthorn97, rand98, otte99, tullmann00}. The ionization state of eDIG is generally compatible with the ionising flux from stellar sources (O and B stars), hardened by photoelectric absorption \citep{hoopes03, rossa03,oey07,haffner09,blanc09}. However, some observations could be more consistent with shocks \citep{shull79}, turbulent mixing layers \citep{slavin93}, or cosmic ray heating \citep{ hartquist86, parker92}. 

Increased \sii /H$\alpha$ and \nii /H$\alpha$ ratios could be explained by increasing the ionization parameter in pure photoionization models. However, the \sii /\nii~  ratio is sensitive to the ionization parameter, and consequently would be required to decrease with disk height \citep{rand98,domgorgen94, blandhawthorn97}.
We show the \sii /\nii~ line ratios for our six galaxies in row G of Figure \ref{maps}. Only spaxels with signal-to-noise ratio (S/N) $>$5 in both lines have been included. 
In the case of Galaxies 4, 5, and 6, the signal-to-noise cut removes most of the extraplanar emission. However, the \sii /\nii~ ratio is not seen to decrease with disk height, which would be required if a change in ionization parameter was causing the increased \sii /H$\alpha$ and \nii /H$\alpha$ observed. In the centre of Galaxy 3 there is a decrease in \sii /\nii~ ratio which corresponds to an increase in ionization parameter. This region is biconical in shape and could be due to shocked outflowing gas. A similar, although less pronounced, decrease in the central line ratios for Galaxy 6 and Galaxy 2 is also seen.
There is an unusual partially resolved region in Galaxy 4 with low \sii /\nii~ ratio (high ionization parameter), spatially coincident with an HII region (as traced by H$\alpha$). This region could correspond to recent star formation and supernovae activity.
The varying \sii /\nii~ ratio in the galaxies in our sample suggests that the enhanced extraplanar emission line ratios are not caused by photoionization alone.

\subsection{Optical diagnostic diagrams}
\begin{figure*}
\includegraphics[width=0.7\linewidth]{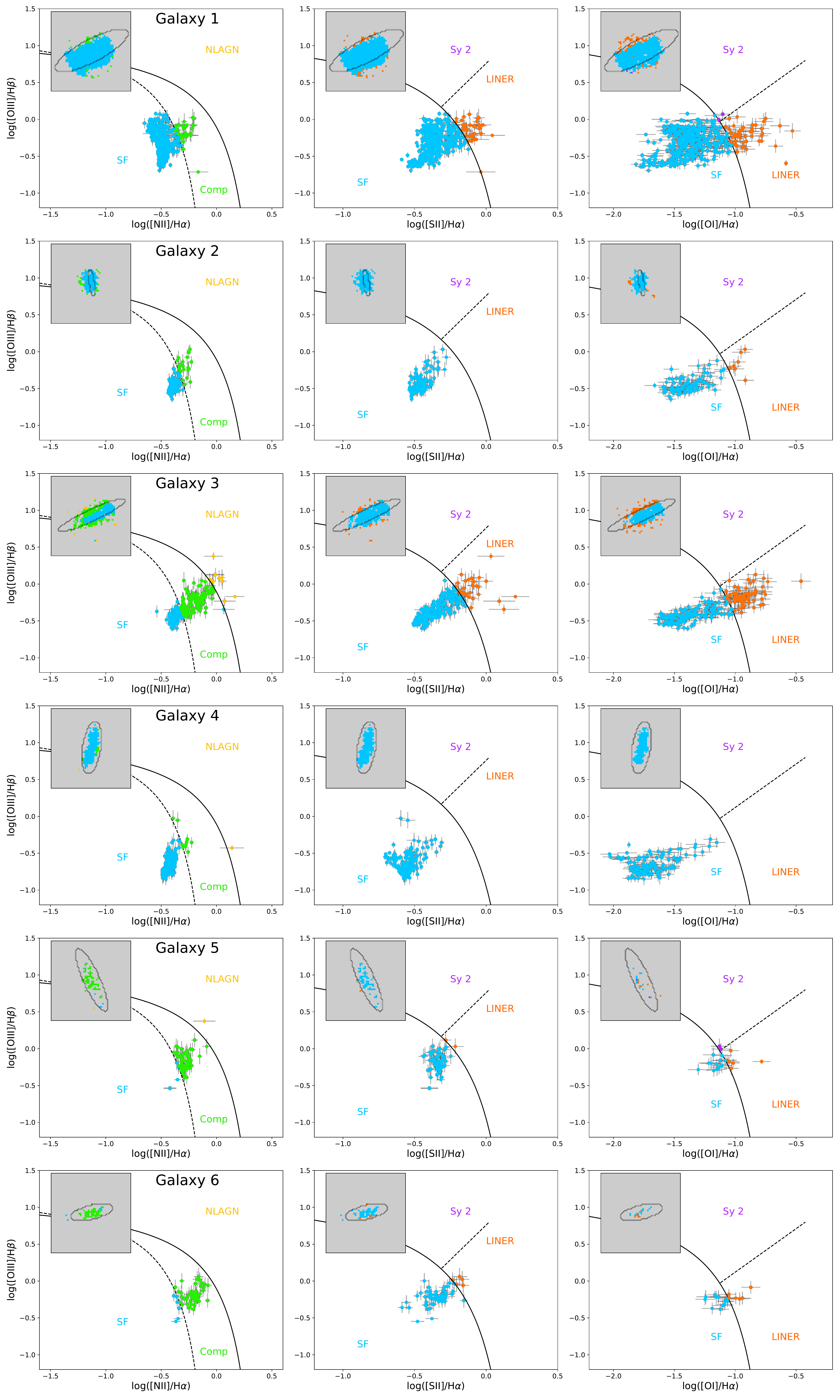}
\caption{Emission line ratio diagnostic diagrams of each spaxel in each galaxy. Black solid curves form an upper limit for star-forming galaxies as derived by Kewley et al. (2001). The dashed line on the \nii~ diagram is the empirical Kauffmann et al. (2003) boundary below which galaxies are classified as star-forming. The dashed lines on the \sii~ and \oi~ diagrams were derived by Kewley et al. (2006) to empirically separate Seyfert 2 galaxies and LINERs.
In the leftmost panel, NLAGN represents narrow emission-line AGN (Seyfert 2 plus LINERs); Comp represents starburst-AGN composites. We apply a minimum S/N cut of 5 in every diagnostic line. The spaxels in each diagram reflect a mix of ionization sources, with spaxels branching from the star-forming regions to the LINER/shocked regions on the diagrams. The insets map the location of the spaxels included in each diagnostic diagram with the same colour coding as their respective diagram. Ellipses show $r_e$ as determined from the r-band photometry.  
}\label{BPTs}
\end{figure*}

Line ratios such as \nii /H$\alpha$, \sii /H$\alpha$, \oi /H$\alpha$ and \oiii /H$\beta$ are sensitive to the hardness of the ionizing radiation field and therefore provide key diagnostics of the ionizing power sources in a galaxy. Diagnostic diagrams using these ratios were first employed by \cite{baldwin81} and \cite{veilleux87} to classify the dominant energy source of a galaxy. These standard optical diagnostic diagrams for the six galaxies in our sample are shown in Figure \ref{BPTs}. We include emission line fluxes with S/N$>$5 in all relevant lines. In all three diagnostic diagrams, emission due to photoionization by H\textsc{II} regions results in line ratios which lie on the lower-left hand side of the diagrams. The observed dashed curve in the \nii~ diagnostic separates purely star forming objects (below the curve) from composite objects containing both star formation and Seyfert 2/LINER-like activity and was derived by \cite{kauffmann03} using the SDSS sample. The dashed line on the \sii /H$\alpha$  and \oi /H$\alpha$ diagnostics represents an empirically derived boundary between Seyfert 2s and LINERs \citep{kewley06}. Shock excitation can cause enhanced line ratios, moving the diagnostic towards the LINER region \citep{rich10,rich11}. Insets in Figure \ref{BPTs} show maps of the classification regions, made using the same colour coding as the data in the diagrams.

All six galaxies have spaxels lying near the purely star forming region, as well as spaxels branching into the composite region. Galaxy 1 shows data consistent with an abundance sequence following the curve from high \oiii /H$\beta$ and low \nii /H$\alpha$, corresponding to low abundances, to regions of lower \oiii /H$\beta$ and higher \nii /H$\alpha$ consistent with higher abundances. The galaxy also shows evidence for a second branch where both \oiii /H$\beta$ and \nii /H$\alpha$ increase together, and also with the disk height as shown in the inset. This second trend can be interpreted as a shock mixing sequence, with shock excitation having an increasing contribution to the emission with increasing line ratios.  Galaxy 4 also shows evidence for both star forming and shock mixing sequences, particularly in the \sii~ diagnostic, as does Galaxy 6. Galaxies 2 and 3 show evidence for a single sequence of shock-mixing. 
Extended LINER-like emission line ratios can be explained by shock models  \citep{allen08, rich10,rich11}, however, they could also be due to other mechanisms such as hot evolved post-AGB stars (see \citealt{Belfiore2016} and references therein). Based on the stacking analysis of \cite{ho16b} (see their figures 12-14), we believe the increase in emission line ratios with galaxy disk height is a true behaviour in galaxies 1-3, not an artefact of our S/N cut.
The composite and LINER-like emission ratios in the galactic winds in galaxies 1-3 results can be explained by wind activity causing shock excitation.

A limiting factor in the optical diagnostic diagrams is the low S/N of the \oiii/H$\beta$ ratio. In Figure \ref{lvdisp}, we compare the \nii/H$\alpha$, \sii/H$\alpha$, and \oi/H$\alpha$ line ratios to their corresponding velocity dispersions. Compared to Figure \ref{BPTs}, more data are shown because we are no longer discarding spaxels with weak H$\beta$ emission that do not meet our S/N criterion.

\subsection{Gas kinematic properties}\label{kinematic}
Gas kinematics were derived at each spaxel using a single Gaussian fit to all emission lines simultaneously with \textsc{lzifu}. Figure \ref{maps} rows E and F show the fitted emission line velocity dispersion and velocity. The velocity dispersion maps show that emission lines tend to broaden away from the disk. This broadening could be due to expanding shock fronts that are too small to spatially resolve or due to two distinct velocity components that we are unable to resolve in our spectra. The extraplanar gas is still rotating in the same sense as the disk.

\cite{ho16} empirically determined criteria to differentiate edge-on galaxies dominated by large-scale winds from  galaxies whose extraplanar emission is dominated by other processes such as diffuse ionized gas or satellite accretion. The criteria involve the asymmetry of the velocity field and the velocity dispersion of the extraplanar gas. Strong disk-halo interactions driven by winds do not have a perfectly symmetric velocity field (e.g. \citealt{shopbell98,sharp10,westmoquette11}). The asymmetry of the velocity field is calculated from the velocity map by reflecting the map over the galaxy major axis. A difference map is constructed by subtracting the reflected velocity from the original velocity map. 

We show this difference map in row G of Figure \ref{maps}.  The standard deviation of the difference maps, weighted by the error on the velocities, gives an estimate of the asymmetry between the two sides of the disk. We only use the difference maps at locations greater than 1 effective radius (shown as a dotted ellipse), where gas entrained in a wind is expected to be an important kinematic component. Following \cite{ho16}, we increase the $r_e$ used by 1 arcsec (denoted as $\tilde{r_e}$) to  reduce the effect of beam smearing. The asymmetry parameter is denoted $\xi$, and is defined in Equations 1 and 2 of \cite{ho16}, repeated here:
\begin{equation}
\xi = \frac{\xi_+ +\xi_- }{2}
\end{equation}
where, 
\begin{equation}
\xi_{+/-} = \underset{r_{+/-} > \tilde{r_{e} }}{\text{std}} \left( \frac{v_{gas} - v_{gas, flipped}}{\sqrt{\text{Err}(v_{gas})^2 + \text{Err}(v_{gas,flipped})^2}}\right)
\end{equation}

The ratio of gas velocity dispersion to rotation is the second parameter used to identify wind galaxies in \cite{ho16}. We utilise the parameter $\eta_{50}$, the ratio of the median velocity dispersion of all spaxels outside 1 $\tilde{r_e}$ with S/N of H$\alpha >5$ to the rotational velocity of the galaxy, $v_{rot}$. 
Because our SAMI observations do not have sufficient spatial coverage to probe the maximum rotational velocity, we have used the stellar mass Tully-Fisher relation of \cite{bell01} to estimate $v_{rot}$ as in \cite{ho16}. 

Values for the two empirical parameters $\xi$ and $\eta_{50}$ are given in Table \ref{params} for our edge-on galaxy sample.  \cite{ho16} empirically classified galaxies as being wind-dominated when $\eta_{50}>0.3$ and $\xi>1.8$. Using these constraints, we  indicate whether or not our galaxies are wind-dominated in Table \ref{params}. We were unable to classify Galaxy 6 due to the high uncertainties  on the velocity dispersions at distances $>1 r_e$, resulting in only two spaxels for analysis ($N_{pix}>$100 is recommended for a robust parameter measurement). Three out of the remaining five galaxies were classified as wind dominated according to the \cite{ho16} criteria. Galaxies 4 and 5 were not classified as wind-dominated, but they lie close to the classification boundary. 
Disk warps and flares are not expected to significantly contribute to the $\eta_{50}$ and $\xi$ parameters in these galaxies \citep{ho16}. Galaxy 2 has the largest asymmetry parameter $\xi =4.6$ which could be enhanced by a possible interaction with a nearby galaxy to the south.
We note that the values of $\eta_{50}$ and $\xi$ were conservatively chosen by \cite{ho16}, based on a sample of 40 galaxies, in order to have a convenient separation into ``wind-dominated'' and ``not wind-dominated'' galaxies. The distribution of galaxy $\eta_{50}$ and $\xi$ values is not bimodal, and the wind classification is not absolute (see Tescari et al (submitted)). Therefore a galaxy with an $\eta_{50}<0.3$ and/or $\xi<1.8$ may still host a weaker wind.  

\begin{table}
\caption[Classifying winds with velocity asymmetry and dispersion parameters.]{Classifying winds with velocity asymmetry ($\xi$) and dispersion parameters ($\eta_{50}$) following \cite{ho16}. Galaxies are classified as ``wind-dominated'' according to \cite{ho16} if $\eta_{50}>0.3$ and $\xi>1.8$. The number of spaxels at radii greater than r$_e$ that meet our S/N criterion is shown in column 2. Unfortunately, there is insufficient data for Galaxy 6 to perform this classification. }\label{params}
\begin{tabular}{|cc|c|c|c|c|}
\hline 
& GAMA ID & N$_{pix}$ &$\eta_{50}$ & $\xi$ & Wind \\ 
\hline 

1 &551202 & 168 & 0.371 & 3.556& Yes \\ 

2& 417678 & 356 & 0.707 & 4.602 &Yes \\ 

3&593680 & 283 & 0.415 & 2.182 & Yes \\ 

4&600030 & 134 & 0.513 & 1.647 &No \\

5&376293 & 221 & 0.455 & 1.43 & No \\ 

6&227607 & 2 & - & - & ? \\ 
\hline
\end{tabular} 

\end{table}

\subsection{Summary of optical diagnostic maps}
\begin{figure*}
\includegraphics[width=0.7\linewidth]{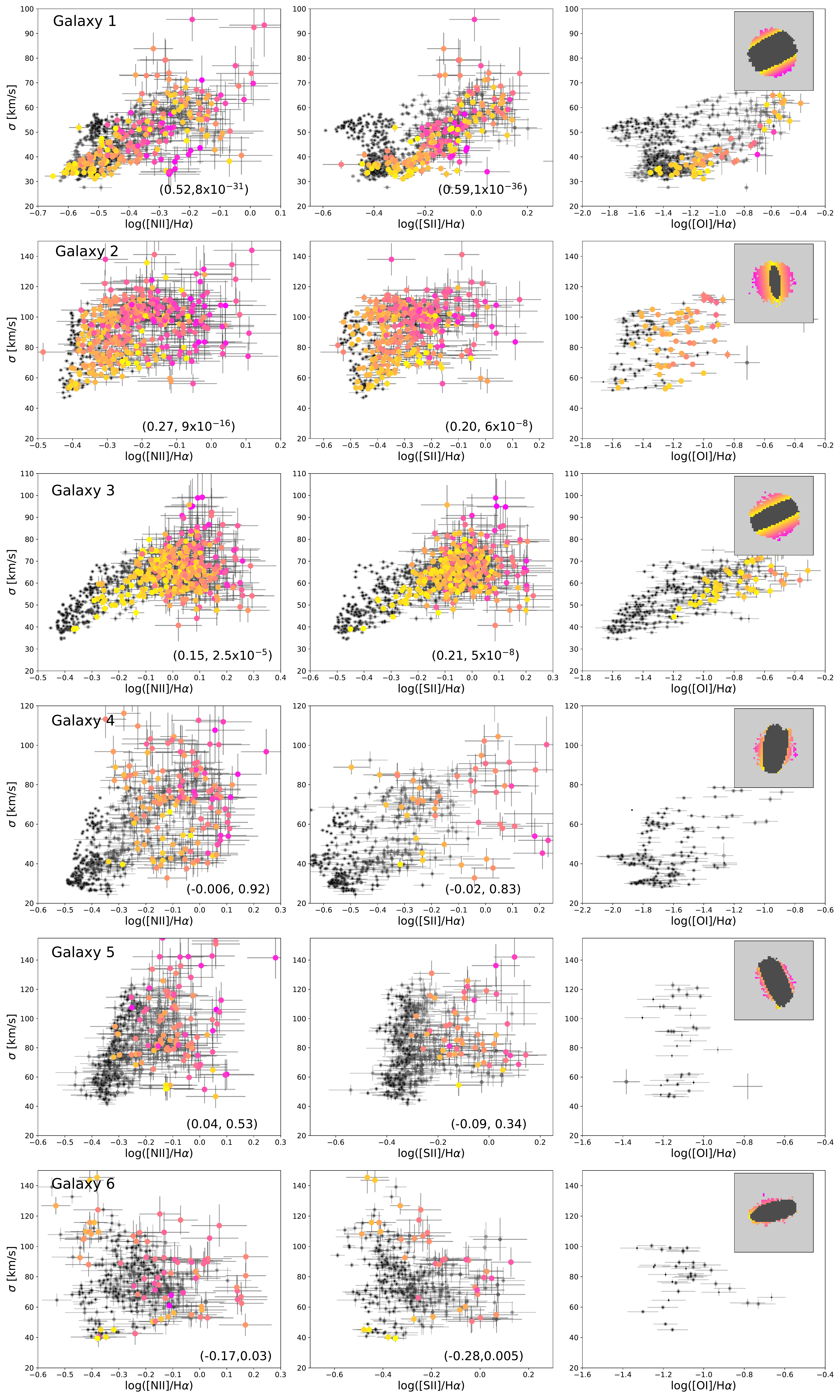}
\caption{Relationship between log(\nii /H$\alpha$), log(\sii /H$\alpha$) and log(\oi /H$\alpha$) and velocity dispersion for the six galaxies. All spaxels with signal-to-noise ratio greater than five in relevant properties H$\alpha$, velocity  dispersion and either \nii~, \sii~, or \oi~ are considered here. Spaxels greater than 1 $\tilde{r_e}$ are colour coded by disk height (as mapped in the inset), which is calculated using the exponential fits to the r-band images by the GAMA team. Spaxels with distances $<\tilde{r_e}$ are considered to belong to the galaxy disk and are coloured dark grey.  Kendall's tau correlation coefficient and significance values were calculated to assess the correspondence between line ratio and velocity dispersion for spaxels outside 1 $\tilde{r_e}$ and they written in the bottom right corner of the \nii /H$\alpha$ and \sii /H$\alpha$ panels. For galaxies 1 to 3, there is a significant positive correlation between line ratio and velocity dispersion in the extra-planar gas. We suggest this is due to shocks, which cause higher line ratios and velocity dispersions in the emission line gas than star formation alone.
}\label{lvdisp}
\end{figure*}

Correlations between between the line ratios \nii/H$\alpha$, \sii/H$\alpha$, and \oi/H$\alpha$ and velocity dispersion have been observed in systems with galactic-wide shock activities \citep{rich11,ho14}.
A positive correlation exists between the line ratios and velocity dispersions of the extraplanar gas for half of our sample.  Figure \ref{lvdisp} shows the relation between emission-line gas $\sigma$, line ratios \nii/H$\alpha$, \sii/H$\alpha$, and \oi/H$\alpha$, and disk-height. The spaxels bifurcate, with the disk and extraplanar regions having different $\sigma$-line ratio distributions. 
We expect shocked gas to have higher emission line ratios and higher velocity dispersion than gas purely excited by star formation. Emission broadened by beam smearing alone should not have higher \nii /H$\alpha$ ratios, because the excitation is still from star formation. Furthermore, beam smearing causes increased velocity dispersions in the centre of a galaxy where the velocity gradient is steepest \citep{green14}.
To focus on the extra-planar gas, we mask out the central r$_e$ (padded by 1 arcsecond) and measure the correspondence between $\sigma$ and the line ratios \nii/H$\alpha$ and \sii/H$\alpha$ using Kendall's tau correlation coefficients ($\rho$) and their significance values (in the range 0 to 1, with smaller numbers indicating higher significance). The scipy.stats package was used for spaxels which have \nii (or \sii), H$\alpha$, and $\sigma$ S/N$>$5 and the values of $\rho$ and significance are included in the bottom right corners of the corresponding panels of Figure \ref{lvdisp}. 

Extraplanar gas in Galaxies 1, 2, and 3, show significant, although weak, positive correlation between the \nii/H$\alpha$ (and \sii/H$\alpha$) line ratio and $\sigma$ ($>5\sigma$), likely due to shock mixing.  

Gas excited by star formation in the disks of galaxies is not expected to show a correlation between emission line ratios and velocity dispersion
\footnote{However, the \nii /H$\alpha$ ratio may show a correlation in the absence of shocks because it is most sensitive to metallicity and therefore is often peaked in the centre of a galaxy. Because the velocity dispersion is often also peaked at the centre of our galaxies, mostly due to beam smearing, a correlation of \nii /H$\alpha$ and $\sigma$ in the disk may be expected.}.
For all galaxies except Galaxy 3 the disk (grey points) does not show a changing line ratio with changing velocity dispersion (i.e. line ratios and $\sigma$ are not correlated). Where there is detectable extraplanar gas, these spaxels tend to show a positive correlation between line ratio, velocity dispersion, and disk height.

\subsection{Star forming properties determined from optical data.}
It has long been known that galaxies with higher SFR surface densities are more likely to host winds \citep{heckman90,heckman15,ho16}. However, it is difficult to conduct inter-sample studies because SFR surface densities are sensitive to the assumptions regarding the SFR tracer and the area under consideration. 

SFRs from H$\alpha$ were derived from SAMI data as follows. We have measured the spectrum of each galaxy in our sample within an elliptical aperture mask of 1 $\tilde{r_e}$. The spectra from the original data cubes were binned (taking spatial covariance into account) and fit with \textsc{lzifu}, as described in Section \ref{specanal}. \textsc{Lzifu} removes the contribution of stellar absorption from the emission lines. The H$\alpha$ emission line flux was corrected for extinction using the Balmer decrement, H$\alpha$/H$\beta$, assuming the expected ratio of 2.86 for Case B recombination and a temperature of T=10$^4$K using the \cite{cardelli89} extinction curve. Extinction-corrected H$\alpha$ fluxes were converted to SFRs using the \cite{kennicutt98} calibration.

The \cite{ho16} sample consists of 40 galaxies, 15 of which were classified as wind-dominated using the kinematic criteria presented in Section \ref{kinematic}.
Using the same area definition as this work, \cite{ho16} found that edge-on wind-dominated galaxies in the SAMI Galaxy Survey have a broad range of  $\Sigma_{SFR}$ $\sim$ [0.001, 0.03) M$_\odot$ yr$^{-1}$ kpc$^{-2}$.  Our sample spans $\Sigma_{SFR}$ of 0.007 to 0.05  M$_\odot$ yr$^{-1}$ kpc$^{-2}$, with galaxies kinematically classified as wind-dominated having the highest 
star formation rate surface densities of our sample ($\Sigma_{SFR} >$0.015 M$_\odot$ yr$^{-1}$ kpc$^{-2}$). Galaxy 4 also has a $\Sigma_{SFR}$ consistent with the wind-dominated galaxies (0.044 M$_\odot$ yr$^{-1}$ kpc$^{-2}$) in our sample, even though its ionized gas kinematics do not show clear signs of winds. We note that our Galaxies 2, 3, and 5 are also in the full sample of \cite{ho16}. Galaxies 1, 4, and 6 were not included in the analysis of \cite{ho16} because they did not meet their stricter inclination criterion. 

Star formation history could play an important role in the detection of winds. Authors such as \cite{ho16}, \cite{sato09}, and \cite{sharp10} find that bursty star formation is preferred over continuous star formation for the driving of winds. 
Using the spectra extracted from an elliptical aperture covering the central $\tilde{r_e}$, we also measured the D$_n$(4000) and EW(H$\delta_A$) indices, which are probes of star formation history \citep{balogh99,kauffmann03}.

D$_n$(4000) is a measurement of the amplitude of the discontinuity at 4000\AA. D$_n$(4000) reflects the mean temperature of stars responsible for the continuum and increases as a function of age. 
We use the definition of \cite{bruzual83}, with the narrower passbands (hence the subscript $n$) proposed by \cite{balogh99} to measure D$_n$(4000) values from our spectra. 
For any stellar type, the value of D$_n$(4000) is also sensitive to stellar metal abundance, being lower for stars of lower metallicity. As a result, the D$_n$(4000) amplitude will be affected by radial abundance gradients of galaxies. By using  spectra from within one effective radius to calculate the D$_n$(4000) values for each galaxy, we should mitigate this problem because local disk galaxies have similar metallicity gradients when calculated in terms of effective radii \citep{roig15}.

We use the method of \cite{worthey97} to measure the equivalent width of the H$\delta_A$ absorption in our 1 $\tilde{r_e}$ spectra. The subscript ``A'' indicates a wide ($\sim 40$\AA) central passband is used. To account for contribution from nebular emission, we include the H$\delta$ emission line in our \textsc{lzifu} fit and subtract the emission-line flux when the line is significant (S/N$>$3). For Galaxy 6 the equivalent width was calculated by hand due to its complex line profile dominated by emission.
A strong H$\delta$ absorption feature is usually interpreted as evidence that a burst of star formation occurred in the last Gyr \citep{couch87}.


Combining the D$_n$(4000) and H$\delta _A$ indices has been employed in a simple diagnostic to distinguish between continuous and burst star formation histories \citep{kauffmann03}.

\begin{figure}
\centering
\includegraphics[width=\linewidth]{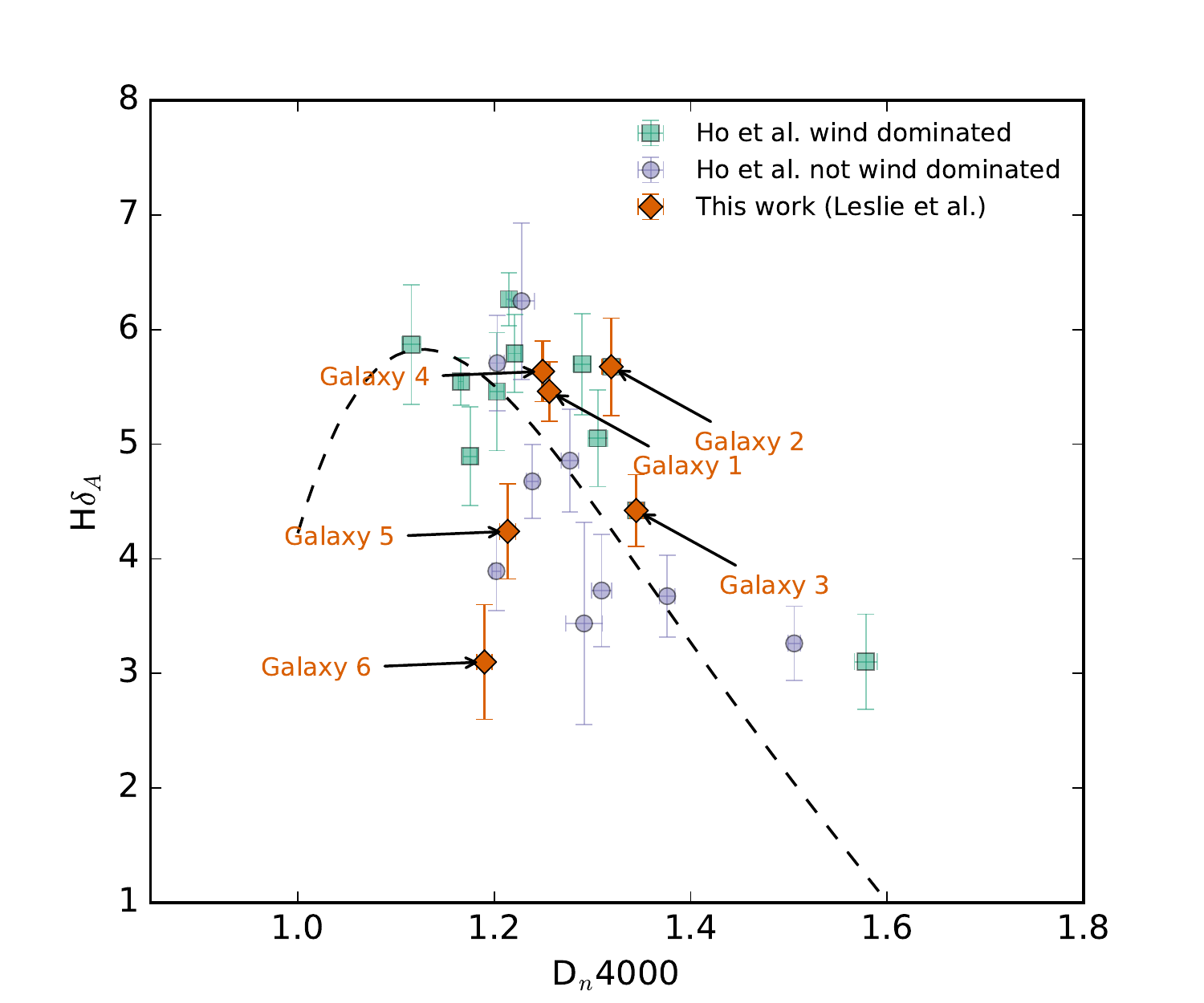}
\caption[Star formation history traced by D$_n$(4000) and H$\delta_A$ of our
galaxies.]{Star formation history traced by D$_n$(4000) and H$\delta_A$ of our
galaxies (orange diamonds). Comparison galaxies with $\log(\Sigma_{SFR})>-2.4$ from \cite{ho16} are also included. The wind-dominated galaxies from \cite{ho16} are shown as green squares and the rest of the comparison
sample as purple circles. The dashed curve indicates the median SDSS relation. Galaxies above this dashed line have `bursty' SFHs.}\label{d4000}
\end{figure}

 \cite{ho16} fit a polynomial to the median relation of the star forming galaxies in SDSS DR 7 on the D$_n$(4000)-H$\delta_A$ plane. Galaxies above the median D$_n$(4000)-H$\delta_A$ relation are more likely to have had a bursty rather than a continuous star formation history. This relation is shown in Figure \ref{d4000} along with our radio-selected sample and the two groups of galaxies in \cite{ho16} with $\Sigma_{SFR}$ matched to our sample ($\Sigma_{SFR} >-2.4$). The FIRST catalogue only contains galaxies with a high surface brightness, biasing our radio-selected sample towards a larger fraction of wind-dominated galaxies. For a fair comparison, we have selected galaxies from \cite{ho16} with $\log(\Sigma_{SFR}) > -2.4$ to compare with our radio-selected sample. This matching results in a sample of 11 wind-dominated galaxies and 9 non-wind dominated galaxies from the original \cite{ho16} sample.

\cite{ho16} found that the wind-dominated galaxies lie on average above the median D$_n$(4000)-H$\delta_A$ relation of the SDSS galaxies, whereas the non-wind-dominated galaxies are distributed both above and below the median. Galaxies with high EW(H$\delta$) for a given D$_n$(4000) value must have been observed a few hundred million years after a burst of star formation - once the emission from the A-star population begins to dominate over the O and B stars (where the Balmer absorption is not as strong). In our sample, Galaxies 1, 2, 3, and 4, with the highest $\Sigma_{SFR}$ all fall above the median relation. This follows the finding of \cite{ho16} that wind-dominated galaxies tend to have a bursty star formation history.

\section{Radio emission}

\begin{figure*}
\centering
\includegraphics[width=\linewidth]{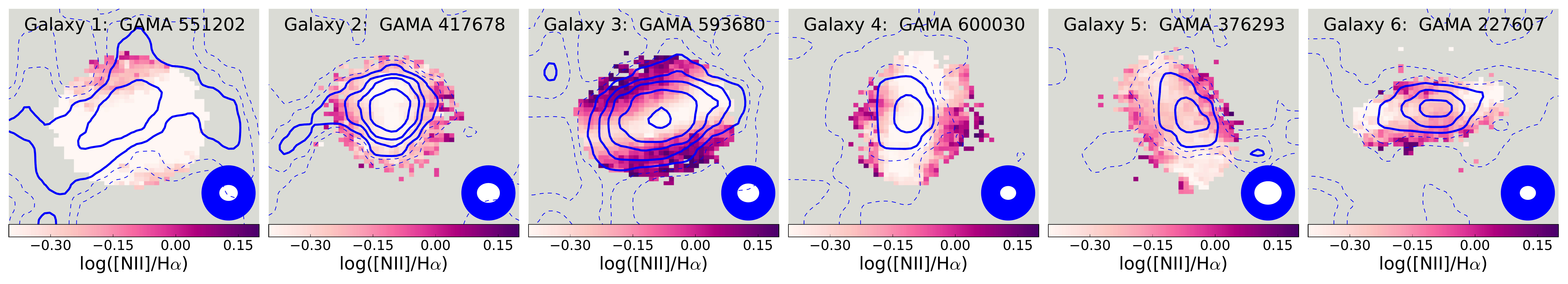}
\includegraphics[width=\linewidth]{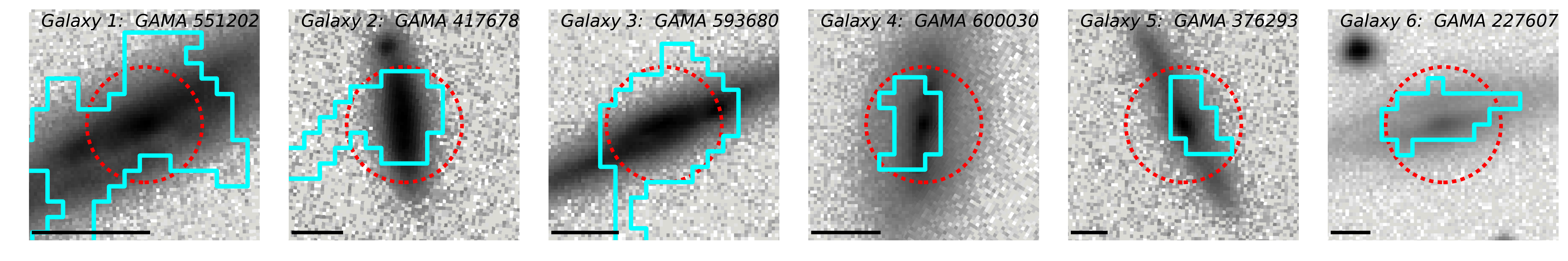}
\caption{Top: Radio contours (blue) overlaid on the \nii /H$\alpha$ map of each galaxy. Solid contours are drawn at  3, 5, 7, 10 and 15 times the rms of the FIRST images (typically 0.15 mJy). Dashed contours are drawn at 1 and 2 times the rms. A white ellipse shows the size of the optical seeing disk and the synthesized beam of the radio observations is indicated by a blue ellipse. The optical images provide data of superior angular resolution. Galaxies 1-3 have wind-dominated kinematics. 
Bottom: The extent of each blob fitted with the \textsc{blobcat} algorithm of \citet{hales12} is outlined in cyan. Blobs encompass regions with S/N$>2$. Each panel is 30 arcsec on the side. The background grey-scale map is the SDSS-r band image. A black bar in the bottom left of the panels indicates 5 kpc at the redshift of each galaxy. The SAMI hexabundle field of view is indicated by a red circle. }\label{overlay}
\end{figure*}

We compare the radio continuum emission measured by the FIRST survey with the optical emission line gas ratios (specifically \nii /H$\alpha$) of our galaxies in Figure \ref{overlay}. The synthesized beam size of the radio images is represented by the blue ellipses at the bottom right of the images. The typical beam size is 5.4$\times$6.4 . Contours have been drawn at 
3, 5, 7, 10 and 15 times the rms of the FIRST images (typically 0.15 mJy; values for individual galaxies are tabulated in Table \ref{sfrs}). 
In Table \ref{sfrs}, we include the extent of the radio emission of the three galaxies with resolved radio emission, estimated as the maximum extent of the detected radio emission above or below the disk.

The radio emission of Galaxies 1, 2, and 3 is extended along the minor axes of the galaxies at signal-to-noise levels of 2-3. These are also the three galaxies classified as ``wind-dominated'' from their kinematic properties. The three wind-dominated galaxies are detected at higher signal-to-noise (by a factor of $\sim$ 3) in the radio images than Galaxies 4, 5, and 6.

The minor axis radio emission is unresolved in the three galaxies not kinematically classified as wind-dominated (Galaxies 4, 5, and 6). The observed size of the minor axis (from the FIRST catalogue) is given in Table \ref{sfrs} as an upper-limit to the vertical extent of the radio emission. Because this measurement is convolved with the radio PSF, the actual size of the radio minor axis could be much smaller than the number given. 
The three galaxies for which we do not report resolved extended minor axis 1.4 GHz emission are at a higher redshift than those in which we do, meaning that they subtend a smaller angle on the sky. The reason that extended radio emission is not detected in these three galaxies could therefore be because the galaxies are not well resolved by the FIRST beam and hence we are unable to resolve any extended emission. In fact, if Galaxy 1 was at the redshift of Galaxies 5 or 6, we would not be able to resolve the extended emission because the beam size is $>5$ kpc, whereas the extended emission only reaches to $\sim$4 kpc above the disk. 

Notes describing the emission of individual objects can be found in the Appendix.


\subsection{Extended radio emission in galaxies with wind signatures}
The galaxies with resolved extended radio emission (Galaxy 1, 2, and 3) are the galaxies kinematically classified as wind-dominated. These galaxies have the largest D$_n$(4000) values, implying their stellar populations are the oldest.  The three galaxies also have radio 1.4 GHz SFRs greater than H$\alpha$ SFRs (Table \ref{sfrs}).
Radio emission traces star formation on timescales $>$10 Myr, whereas H$\alpha$ emission traces star formation on timescales $<$10 Myr. An older stellar population rather than a young ($<10$ Myr) population is more likely to be detected and extended in the radio because it takes time for the radio emission to be produced in SNe and for the CRs to be transported to the large scale heights. Larger radio than H$\alpha$ SFRs in Galaxies 1, 2, an 3 are therefore consistent with a bursty star formation history. 

To further investigate the extended radio emission in our sample we use the \textsc{blobcat} software \citep{hales12}, which uses the flood fill algorithm to detect and catalogue ``blobs'' that represent sources in a 2D astronomical image. In particular, \textsc{blobcat} has been designed to analyse the Stokes I intensity and linear polarization of radio-wavelength images. 
\textsc{Blobcat} was used to analyse 30 by 30 arcsec cut-out FIRST images of each of our six objects. We required a detection S/N (signal per pixel / rms noise in the FIRST map) of at least three\footnote{\textsc{Blobcat} is optimised for S/N$>$5 detection threshold. However, to detect Galaxy 4, 5, and 6, a S/N detection threshold of 3 was required.} and blobs include connecting pixels down to 2 sigma flux levels. The rms values were taken from the FIRST catalogue, which gives local rms values computed by combining the measured noise from all grid pointing images contributing to the co-added map position.  We set the CLEAN bias correction to 0.25 mJy as suggested by \cite{becker95}. 

The \textsc{blobcat} region of each source is shown overlaid on SDSS $r$ band images in the bottom row of Figure \ref{overlay}.

\textsc{Blobcat} returns a parameter R$_{EST}$ that can be used to identify sources with non-Gaussian complex morphologies. R$_{EST}$ is the ratio of the area of the detected blob to the area covered by an unresolved Gaussian blob with the same peak surface brightness (with local bandwidth smearing taken into account). This method assumes an elliptical Gaussian point spread function, whose parameters are given in the header of the FIRST image. The further the value from 1, the more non-Gaussian the object is and the more extended. We list the values of R$_{EST}$ obtained from \textsc{blobcat} in Table \ref{sfrs}.
R$_{EST}$ is largest for the three galaxies classified as wind-dominated, suggesting that R$_{EST}$ could be used to identify galaxies with a potential wind in radio images. However, Galaxy 2 has an R$_{EST}$ (1.59) similar to the non-wind dominated Galaxies 4-6 (mean 1.16).

\subsection{Infrared-Radio Correlation}

We calculate the total infrared luminosities (TIR; 8-1000 $\mu$m) of our sample using the calibration of \cite{galametz13} involving 24, 100 and 160 $\mu$m data, but drawing on the WISE 22$\mu$m emission in place of the unavailable 24 $\mu$m Spitzer data. The 100 and 160$\mu$m data are from the Herschel-ATLAS first data release \citep{bourne16,valiante16}.

The TIR-radio correlation is parametrized by $q$, 
\begin{equation}
q = \log\left(\frac{L_{TIR}}{3.75\times 10^{12} \text{W}}\right) - \log\left(\frac{L_{1.4GHz}}{\text{W  Hz}^{-1}}\right). 
\label{q}
\end{equation}
\cite{bell03} found for a sample of local galaxies an average $q = 2.64\pm 0.26$. The $q$ values of our sample are included in Table \ref{sfrs}.
We find that the three galaxies classified as wind-dominated from their kinematics have the lowest $q$ values (mean value 2.54$\pm$0.16 compared to a mean of 2.73$\pm$0.04 for Galaxies 4-6). Although the $q$ values are consistent with star-forming galaxies, a lower $q$-value means that the non-thermal radio synchrotron emission is dominant over the thermal infrared emission.

\begin{table*}
\caption{Derived properties of the six galaxies in our sample. Radio SFRs are measured using global 1.4 GHz fluxes from FIRST following \citep{mauch07}. 
The star formation rate surface densities, SFR$_{r_e}/(\pi r_e^2)$, is measured using H$\alpha$ emission-line flux from within 1 effective radius once corrected for extinction using the Balmer decrement.
D$_n$(4000) and H$\delta_A$ are calculated from the SAMI spectra within 1 $r_e$. 
q$_{TIR}$ is the ratio of infrared to radio luminosity defined in Equation \ref{q}.
The second last column shows the `radio extent'; this is the scale height at which we detect resolved minor axis radio emission in the FIRST images. For galaxies without resolved minor axis emission, we give an upper limit. Sizes are converted to kpc based on the redshift of each galaxy. R$_{EST}$ is the ratio of the area covered by each source to the area of an unresolved Gaussian source.  }

\label{sfrs}
\begin{tabular}{|cc|ccc|ccccc|c|c|c|}
\hline
&GAMA ID & FIRST flux & FIRST rms & SFR$_{1.4GHz}$ & log($\Sigma_{SFR,H\alpha}$) & D$_n$(4000) & H$\delta_A$ & $q_{TIR}$ & Radio extent |z| & R$_{EST}$\\
&& mJy & mJy/beam & M$_\odot$yr$^{-1}$ & log(M$_\odot$yr$^{-1}$kpc$^{-2}$)& & \AA & & kpc & \\
\hline

1&551202 & 7.096 & 0.1558 &7.1 & -1.52 & 1.26 & 5.5 & 2.56$\pm$0.03 & $\sim$4 & 8.93 \\
2&417678 & 2.87 & 0.1480 & 2.9& -1.29 & 1.32 & 5.7 & 2.64$\pm$0.04  & $\sim$8 & 1.59\\
3&593680 & 5.676 & 0.1453 &5.7 & -1.81 & 1.34 & 4.4 & 2.63$\pm$0.03 & $\sim$ 5 & 2.42\\
4&600030 & 1.982 & 0.1422 &2.0 & -1.35 & 1.25 & 5.6& 2.69$\pm$0.06 & $<$3.6 & 1.21\\
5&376293 & 1.260 & 0.1540 &1.3 & -2.03 & 1.21 & 4.2 & 2.77$\pm$0.09 & $<$5.2 & 0.91\\
6&227607 & 2.106 & 0.1535 & 2.1& -2.17 & 1.19 & 2.0 & 2.72$\pm$0.04 & $<$5.6 & 1.36\\
\hline
\end{tabular}
\end{table*}

\section{Discussion}\label{discussion}

We have considered different indicators for wind activity in our galaxies from optical gas emission and kinematic diagnostics to radio morphology. Table \ref{evidence} summarises the evidence for and against dominant wind activity in each galaxy.
Emission at 1.4 GHz is produced by both thermal emission and synchrotron processes. 1.4 GHz emission is expected in both thermal and CR driven winds. Supernovae both heat the gas and accelerate the CRs.  
In this section we discuss how galactic winds potentially driven by a combination of CR pressure and SNe activity can explain our observations. We then discuss future observations that could further constrain our interpretations.



Galaxies with a starburst driven wind have ionisation properties consistent with shock processes. 
\cite{dopita96} found that $\sim$50\% of mechanical energy from the wind is converted into energy capable of ionizing gas. If the SAMI data are tracking material that has been entrained by a wind, then the emission lines with high line ratios are likely to be excited by shocks. A correlation between the FWHM of emission lines and shock velocity can occur when radiative shocks move into denser clouds that have a fractal distribution \citep{dopita12}. Further, a one-to-one correlation can be explained by the observation of two shocks moving in opposite directions \citep{dopita12}. 
Our data are unable to resolve individual shock fronts, so shocks could be propagating in any direction within an individual spaxel. As a result, one might expect the velocity dispersions would be related to the wind velocities in our sample. The maximum velocity dispersions measured ($\sim 150$ km s$^{-1}$) for the six galaxies in our sample are smaller than the velocity dispersions measured for galactic scale starburst-driven super-winds. However, the velocity dispersions of the six galaxies are consistent with velocity dispersions measured in the wind-dominated galaxies of \cite{ho16}.

Cosmic ray-driven winds, as described in \cite{booth13}, result in lower wind velocities than thermal feedback models alone. For example, simulations by \cite{booth13} predict an outflow velocity for a Milky Way-like galaxy of $\sim$200 km s$^{-1}$ rather than the $\sim$1000 km s$^{-1}$ predicted by thermal feedback models. \cite{girichidis16} also found that thermally driven-winds are hot, mostly made of ionized hydrogen and have low densities ($\sim$10$^{-27}$ g cm$^{-3}$), whereas CR-driven winds are one to two orders of magnitude denser ($\sim$10$^{-26}$- 10$^{-25}$ g cm$^{-3}$), are warm (10$^4$K) and are composed of a mixture of atomic and molecular hydrogen. The gentle acceleration of the ISM that occurs in CR-driven outflow simulations results in multiphase winds, which include a cool component generally not present in thermally-driven winds \citep{girichidis16}. 
Modelling the wind geometry to obtain a wind velocity and comparing observations of molecular and atomic hydrogen gas with the predictions of \cite{girichidis16} and \cite{simpson16} will help understand the relative importance of CRs as a wind-driving mechanism. 

In localised regions of active star formation, superbubbles can locally inject a substantial amount of energy into the surroundings, allowing hot gas to reach considerable heights \citep{deavillez04}. When considering the effect of CRs, which help to push the plasma against the gravitational pull of the galaxy, even a late-type spiral can achieve a steady large-scale wind with a global mass loss rate of $\sim$0.3M$_\odot$ yr$^{-1}$ and a low velocity of $\sim$10 km s$^{-1}$ \citep{breitschwerdt93}. 
If stellar winds and supernovae are responsible for the large amounts of extraplanar gas, with chimneys being the main mode of CR and gas transport, then one might expect to see a correlation of multiple ISM tracers as they follow individual filaments. \cite{dettmar92} found such a correlation between filaments of H$\alpha$ and radio continuum emission features in NGC 5775. Future sensitive high resolution radio and optical imaging will help constrain any spatial correlation on a larger sample. We observe a correlation between SFR surface density and the extent of the minor axis radio emission. Such a correlation could imply that a large amount of star formation activity is required to blow a hole in the disk.

Synchrotron protrusions from edge-on disks \citep{reuter91,seaquist91} and ordered halo magnetic fields \citep{hummel90, hummel91} can also be indicative of winds. The radio emission that extends slightly in the direction of the minor axes of Galaxy 1 and 3 appear similar to the polarized intensity of galaxies such as NGC 5775, which trace the magnetic field and a possible bi-conical outflow \citep{tullmann00,duric88,soida11}
 Some synchrotron protrusions in Galaxy 1, 2, and 3 are observed. Although these protrusions hint that CRs are playing an important role in the disk-halo interaction, we do not have radio data of sufficient resolution or sensitivity to answer questions about the nature of the CRs and what is driving the winds. Electrons lose energy as they propagate through synchrotron radiation and interactions with ambient photons via the inverse Compton process \citep{collins00} changing the spectral index of the non-thermal emission. As such, spectral ageing (e.g. \citealt{heesen09}) could constrain the mode of cosmic ray transport \citep{uhlig12}. Higher resolution radio imaging of the disk at multiple frequencies will help to separate the synchrotron from thermal emission and determine the origin of the cosmic ray emission. 

\begin{table}
\caption{Evidence for wind activity}\label{evidence}

\begin{tabular}{cc|cccc}
\hline
 &GAMA ID & Line ratios & Kinematics & SFH & Radio size\\
\hline

1&551202 & \checkmark & \checkmark & \checkmark & \checkmark \\
2&417678 & \checkmark & \checkmark & \checkmark & \checkmark \\
3&593680 & \checkmark & \checkmark & \checkmark & \checkmark \\
4&600030 & $\times$ & $\times$ & \checkmark & $\times$ \\
5&376293 & $\times$ & $\times$ & $\times$ & $\times$ \\
6&227607 & $\times$ & - & $\times$ & $\times$ \\
\hline 
\end{tabular}
\end{table}
\section{Conclusion}

We have studied six edge-on star-forming galaxies in the SAMI Galaxy Survey that are detected in the 1.4 GHz VLA FIRST survey. These six galaxies are candidates for studying cosmic ray-driven outflows in star-forming galaxies. 
Shock-like emission line ratios are found in all six objects. Significant positive correlation between velocity dispersion, \nii/H$\alpha$ ratio, and disk height indicative of shocks is found for three out of six galaxies. Galaxies 1, 2, and 3 are classified as wind-dominated using the gas kinematic criteria of \cite{ho16}. The extraplanar gas, interpreted as the outflowing component, is observed to be rotating in the same sense as the disk. This co-rotation is not surprising as the outflowing material is expected to entrain part of the rotating disk-material. We discuss the details of each object and speculate on likely explanations for our observations in the Appendix.

Galaxies in our sample with extended radio emission have wind-like signatures. The three galaxies in our sample with extended radio emission have bursty star formation histories, high specific star formation rates and high star formation rate surface densities compared to the galaxies without detectable protrusions. Because these properties are indicative of starburst-driven winds, our observations imply that star formation, cosmic ray emission, and outflows are intimately related. 

Work is still required to understand winds in star-forming galaxies both theoretically and observationally. CRs could help explain the presence of winds and high mass-loading factors in some galaxies. However, the relative importance of CR processes is still unconstrained because extraplanar gas is diffuse and difficult to observe. The comparison of IFU data to future deep radio observations with high angular resolution and covering a range of frequencies will allow us to observationally determine how common CR-driven winds are in the Universe and their relative importance in transporting material into galaxy halos.

\section*{Acknowledgements}
This research was conducted by the Australian Research Council Centre of Excellence for All-sky Astrophysics (CAASTRO), through project number CE110001020.

The SAMI Galaxy Survey is based on observations made at the Anglo-Australian Telescope. The Sydney-AAO Multi-object Integral field spectrograph (SAMI) was developed jointly by the University of Sydney and the Australian Astronomical Observatory. The SAMI input catalogue is based on data taken from the Sloan Digital Sky Survey, the GAMA Survey and the VST ATLAS Survey. The SAMI Galaxy Survey is funded CAASTRO and other participating institutions. The SAMI Galaxy Survey website is http://sami-survey.org/.

GAMA is a joint European-Australasian project based around a spectroscopic campaign using the Anglo-Australian Telescope. The GAMA input catalogue is based on data taken from the Sloan Digital Sky Survey and the UKIRT Infrared Deep Sky Survey. Complementary imaging of the GAMA regions is being obtained by a number of independent survey programs including GALEX MIS, VST KiDS, VISTA VIKING, WISE, Herschel-ATLAS, GMRT and ASKAP providing UV to radio coverage. GAMA is funded by the STFC (UK), the ARC (Australia), the AAO, and the participating institutions. The GAMA website is http://www.gama-survey.org/ .

Support for AMM is provided by NASA through Hubble Fellowship grant \#HST-HF2-51377 awarded by the Space Telescope Science Institute, which is operated by the Association of Universities for Research in Astronomy, Inc., for NASA, under contract NAS5-26555. 
JTA acknowledges the award of a SIEF John Stocker Fellowship. 
MSO. acknowledges the funding support from the Australian Research Council through a Future Fellowship Fellowship (FT140100255).
SB acknowledges the funding support from the Australian Research Council through a Future Fellowship (FT140101166).

\bibliographystyle{mnras}
\bibliography{mybib}

\appendix

\section{Notes on individual objects}


\subsection{GAMA 551202}

Although only the central regions ($\pm$ 2.5 kpc) of GAMA 551202 are probed with SAMI, evidence for increased velocity dispersion and emission line ratios above and below the disk is seen, particularly north of the disk away from the dust lane. GAMA 551202 is the least massive galaxy in the sample, but its  D$_n$(4000) and H$\delta_A$ values (H$\delta_A\sim$ 0.47 \AA~ above the median relation) indicate that a burst of star formation could have occurred in the past 1 Gyr. 
GAMA 551202 is the nearest galaxy in our sample and its large angular size allows us to resolve the extraplanar radio emission. This galaxy has extended radio emission that is offset from the optical brightness profile (position angles differ by 30 degrees). Fingers of radio emission can be seen perpendicular to the disk of GAMA 551202 out to 4 kpc. It is possible that the extended radio emission is evidence for CR-driven winds (or at least the presence of a galactic wind or chimney). The optical emission line ratios for this galaxy show that both star formation and shocks are likely contributing to the emission. It is possible that a low velocity CR-driven wind ($\ll$ 100 km s$^{-1}$) is present in GAMA 551202, but more observations are required to confirm this.

\subsection{GAMA 417678}
GAMA 417678 is the most extreme galaxy in the sample, having the highest optical SFR and $\Sigma_{SFR}$ as well as the highest $\eta_{50}$ and $\xi$ kinematic parameters. GAMA 417678 is likely to have experienced a recent burst of star formation because it lies well above the median SDSS relation in the D$_n$(4000)-H$\delta_A$ plane. 
GAMA 417678 is in a close pair (with GAMA 417676, M$_* = 10^{9.98}$M$_\odot$).
Given the close proximity to another massive galaxy, a recent burst of star formation resulting from an interaction is likely for GAMA 417678. An interaction could also be the reason for the relatively large velocity dispersions (of $\sim$100 km s$^{-1}$) observed in the galaxy disk. One of the most noticeable features of the radio emission from GAMA 417678 is a strong ($>3\times$rms)  tail extending $\sim$ 7.5 kpc to the south-east. This tail is not in the direction of any neighbouring galaxies, so it is possible that the tail may be a result of a starburst-driven galactic wind which has blown a hole in the disk, allowing CRs to be advected into the halo. However, we cannot rule out the possibility that the tail is a tidal feature.
The tail could also be a weak radio jet. If this were a jet, GAMA 417678 would be a rare galaxy that has high SF activity and a large scale radio jet: Most local radio-galaxies are early type objects with red colours \citep{smolcic09}.

\subsection{GAMA 593680}

GAMA 593680 shows extended radio emission reaching to heights of $\sim$5 kpc above the disk. Furthermore, the \sii /\nii~ line ratio map shows evidence of a bipolar outflow, likely a result of a supernova-driven wind. Like the other wind-dominated galaxies, GAMA 593680 lies above the median D$_n$(4000)-H$\delta_A$ relation of SDSS star forming galaxies (by 0.47 \AA). GAMA 593680 has the largest D$_n$(4000) value, implying it has the oldest stellar populations. The strong increase of emission line ratio with disk height (\sii /H$\alpha$ increases by $\sim$0.2 dex parsec$^{-1}$) argues against the line ratios being caused by AGN activity and supports the idea that shocked gas is being expelled from the galaxy in a starburst-driven galactic wind, likely aided by cosmic ray pressure. 
The radio emission is strongly centred on the nucleus of GAMA 593680. This could be due to the high concentration of on-going star formation in the centre of the galaxy. GAMA 593680 has no obvious nearby companions, but it is the centre of a small group of small galaxies (GAMA group catalogue; \citealt{robotham11}). Although the extraplanar radio emission is close to the noise level, extended emission similar to the structure seen in GAMA 551202 can be inferred from the FIRST image out to disk heights of $\sim$ 5 kpc. 

\subsection{GAMA 600030}
Clumps of H$\alpha$ emission are seen to the north and south of the nucleus of GAMA 600030, which are not present in the optical continuum nor radio continuum emission. 

GAMA 600030 shares some properties with the galaxies that have extended radio emission - it has high sSFR and high $\Sigma_{SFR}$. On the other hand, its radio emission is only 40\% of that expected from the H$\alpha$ SFR and the emission was not spatially resolved in the FIRST images. It is possible that this galaxy does not have a strong CR-pressure component, due to a lack of strong magnetic field. The large star formation rate surface density (0.044 M$_\odot$ yr$^{-1}$ kpc$^{-2}$) could drive a thermal outflow, raising the velocity dispersion (from 50 km s$^{-1}$ to 150 km s$^{-1}$) while producing shocks as the outflowing gas collides with its surroundings thereby raising the emission line ratios.

GAMA 600030 has the second highest SFR and has a $\sigma$ map very similar to that of GAMA 417678. GAMA 600030 could be wind dominated. GAMA 600030 lies above (0.58\AA) the median SDSS D$_n$(4000) - H$\delta_A$ relation. It is therefore plausible that this galaxy is experiencing a starburst-driven (radiation pressure and/or cosmic ray-driven) wind. There is an interesting region of low \sii/\nii ratio south of the nucleus of this galaxy. This region coincides with a knot of bright H$\alpha$ emission and could be a signature of a wind-blown super-bubble, although high resolution, multi-wavelength observations are required to test this hypothesis.

\subsection{GAMA 376293}
GAMA 376293 is an isolated galaxy with a knot of star formation on the south-west side of its disk seen in both the H$\alpha$ and optical continuum images. This region also corresponds to a region where there is a possible extraplanar clump of radio emission at 2$\times$ the background rms. 
GAMA 376293 lies on the boundary of being  classified as wind-dominated or not wind-dominated. Based on the gas kinematic classification of \cite{ho16}, GAMA 376293 has the high ratio of velocity dispersion-to-rotational velocity expected for a wind-dominated galaxy but its velocity field is too symmetrical. 
Given that galaxies with CR-driven winds have not been well studied observationally, it is possible that they result in more symmetrical velocity fields than thermal winds. The velocity dispersion of the emission line gas of GAMA 376293 is largest in the nucleus of the disk and above and below the disk, where the emission line ratios are enhanced. This spatial correlation is in support of a galactic wind in GAMA 376293.  The central spectrum of GAMA 376293 has H$\delta_A$ values lower than expected for its D$_n$(4000) given the median SDSS relation. \cite{ho16} found that galaxies below the median relation tend to not be wind-dominated. Therefore, the low H$\delta_A$ value may be further evidence against a galactic wind. 

\subsection{GAMA 227607}

GAMA 227607 is the most massive galaxy in our sample with a clear bulge and dust lane. GAMA 227607 is the most difficult candidate in which to find a wind because of the low S/N spectra obtained with SAMI away from the disk. No evidence is seen for extended radio continuum emission (down to 0.3 mJy), expected if CRs were aiding the driving of gas into the halo. The galaxy does show a bi-conical \nii /H$\alpha$ emission line ratio map and composite spectra which could reflect a contribution from shocks. 
The [OIII]/H$\beta$ ratio is not typical of Seyfert emission, but is consistent with slow shocks (e.g \citealt{rich11}). 
GAMA 227607 is in a group with another star-forming disk galaxy (GAMA 227619) with stellar mass 10$^{10.59}$M$_\odot$, located south-east of GAMA 227607. An interaction with GAMA 227619 could explain the region of high velocity dispersion gas ($\sigma$=140 km s$^{-1}$) to the west of the disk in GAMA 227607. The radio emission of GAMA 227607 is concentrated along the galaxy disk and is likely to be associated with star formation activity. It is possible that GAMA 227607 hosts a weak galactic wind, but it is perhaps more likely that the enhanced emission line ratios seen are a result of diffuse ionized gas.

\label{lastpage}
\end{document}